\documentclass{article}
\pdfoutput=1

\usepackage{trcover}
\usepackage{makeidx,graphicx,booktabs,xspace,url,listings,color,multicol,pifont,multirow,textcomp,subfig,setspace,paralist}

\lstnewenvironment{Bench}[2]
{\lstset{ %
showlines=true,
xleftmargin=5ex,
firstnumber=1,
language=C++,                % choose the language of the code
basicstyle=\scriptsize,       % the size of the fonts that are used for the code
numbers=left,                   % where to put the line-numbers
numberstyle=\tiny,      % the size of the fonts that are used for the line-numbers
stepnumber=2,                   % the step between two line-numbers.
numbersep=3pt,                  % how far the line-numbers are from the code
showspaces=false,               % show spaces adding particular underscores
showstringspaces=false,         % underline spaces within strings
showtabs=false,                 % show tabs within strings adding particular underscores
tabsize=2,                      % sets default tabsize to 2 spaces
captionpos=b,                   % sets the caption-position to bottom
breaklines=true,                % sets automatic line breaking
breakatwhitespace=false,        % sets if automatic breaks should only happen at whitespace
escapeinside={\%*}{*)},          % if you want to add a comment within your code
mathescape=true,
columns=flexible,
morekeywords={main,add_workers,add_emitter,wrap_around,
run_and_wait_end,ff_send_out},
#1%, label={code:#2}
}}{}

\lstnewenvironment{Bench2}[2]
{\lstset{ %
xleftmargin=5ex,
linewidth=1.1\linewidth,
firstnumber=1,
language=C++,                % choose the language of the code
multicols=2,
basicstyle=\scriptsize,       % the size of the fonts that are used for the code
numbers=left,                   % where to put the line-numbers
numberstyle=\tiny,      % the size of the fonts that are used for the line-numbers
stepnumber=2,                   % the step between two line-numbers.
numbersep=3pt,                  % how far the line-numbers are from the code
showspaces=false,               % show spaces adding particular underscores
showstringspaces=false,         % underline spaces within strings
showtabs=false,                 % show tabs within strings adding particular underscores
tabsize=2,                      % sets default tabsize to 2 spaces
captionpos=b,                   % sets the caption-position to bottom
breaklines=true,                % sets automatic line breaking
breakatwhitespace=false,        % sets if automatic breaks should only happen at whitespace
escapeinside={\%*}{*)},          % if you want to add a comment within your code
mathescape=true, columns=flexible, morekeywords={main,add_workers,add_emitter,wrap_around,run_and_wait_end,ff_send_out},
#1%, label={code:#2}
}}{}

\newcommand {\lines}[2]{\textsc{\small \S #1.\oldstylenums{#2}}}
\newcommand {\linesint}[3]{\textsc{\small \S #1.\oldstylenums{#2}-\oldstylenums{#3}}}
\newcommand {\linespair}[3]{\textsc{\small \S #1.\oldstylenums{#2},\S #1.\oldstylenums{#3}}}

\newcommand {\ff}{FastFlow\xspace}
\newcommand {\yadt}{YaDT\xspace}
\newcommand {\yadtff}{YaDT-FF\xspace}

\title{Porting Decision Tree Algorithms to Multicore using \ff}
\trnum{TR-10-11}
\author{Marco Aldinucci\thanks{Computer Science Department, University
    of Torino, Italy. Email: adinuc@di.unito.it} \and Salvatore Ruggieri
  \and Massimo Torquati}
\date{May 18th, 2010}

\begin{document}
\maketitle
\begin{abstract}
The whole computer hardware industry embraced multicores. For these machines, the extreme
optimisation of sequential algorithms is no longer sufficient to squeeze the real machine power,
which can be only exploited via thread-level parallelism. Decision tree algorithms exhibit natural
concurrency that makes them suitable to be parallelised. This paper presents an approach for
\emph{easy-yet-efficient} porting of an implementation of the C4.5 algorithm on multicores. The
parallel porting requires minimal changes to the
original sequential code, and it is able to exploit up to $7\times$ speedup on an Intel dual-quad core machine.
\end{abstract}

\section{Introduction}

%Historically, parallel computing has been considered to be \emph{the
%  high-end of computing}, and has been used to model difficult
%scientific and engineering problems found in the real world. Today,
%commercial applications provide an equal or greater driving force in
%the development of faster computers. These applications require the
%processing of large amounts of data, often in sophisticated ways: data
%classification and mining are bright examples.

Computing hardware has evolved to sustain an insatiable demand for high-end performances along two
basic ways. On the one hand, the increase of clock frequency and the exploitation of
instruction-level parallelism boosted the computing power of the single processor. On the other
hand, many processors have been arranged in multi-processors, multi-computers, and networks of
geographically distributed machines. This latter solution exhibits a superior peak  performance,
but it incurs in significant software development costs. In the last two decades, the parallel
computing research community aimed at designing languages and tools to support the seamless porting
of applications and the tuning of performances \cite{openMP,BlumofeJoKu96,van:assist:02,streamIt}.
These languages, apart from few exceptions that also focus on code portability
\cite{openMP,van:assist:02}, require a redesign of the application logic in an explicitly parallel
language or model.

Up to now, clock speed and algorithmic improvements have exhibited a better performance/cost
trade-off than application redesign, being the possibility to preserve the existing code its most
important component. Data mining is not an exception in this regard. By surveying the papers in the
main scientific conferences and journals, there is a diminishing number of proposals for parallel
implementations of data mining algorithms in the last few years. After all, only a small percentage
of data analysis projects can afford the cost of buying (and maintaining) a parallel machine and a
data mining software capable of exploiting it. In most cases, data reduction techniques (such as
sampling, aggregation, feature selection) can mitigate the problem while
waiting the advancement in memory and computational power of low-cost workstations. %desktop PCs.

Nowadays, however, this vision should be reinterpreted. After years of continual improvement of
single core chips trying to increase instruction-level parallelism, hardware manufacturers realised
that the effort required for further improvements is no longer worth the benefits eventually
achieved. Microprocessor vendors have shifted their attention to thread-level parallelism  by
designing chips with multiple internal cores, known as Multicore or Chip Multiprocessors. However,
this process does not always translate into greater CPU performance: multicore are small-scale but
full-fledged parallel machines and they retain many of their usage
problems. In particular, sequential code will get no performance benefits from them. A workstation %desktop PCs
equipped with a quad-core CPU but running sequential code is wasting $3/4$ of its computational
power. Developers, including data miners, are then facing the challenge of achieving a trade-off
between performance and human productivity (total cost and time to solution) in developing and
porting applications to multicore. Parallel software engineering engaged this challenge trying to
design tools, in the form of high-level sequential language extensions and coding patterns, aiming
at simplifying the porting of sequential codes while guaranteeing the efficient exploitation of
concurrency \cite{openMP,BlumofeJoKu96,van:assist:02,patterson:cacm:09}.

This paper focuses on achieving this trade-off on a case study by adopting a methodology for
the \emph{easy-yet-efficient} porting of an implementation of the C4.5 decision tree induction
algorithm \cite{Qui93} onto multicore machines. We consider the \yadt (Yet another Decision Tree
builder) \cite{yadt} implementation of C4.5, which is a from-scratch and efficient C++ version of
the well-known Quinlan's entropy-based algorithm. \yadt is the result of several data structure
re-design and algorithmic improvements over Efficient C4.5 \cite{ec45}, which is in turn is a patch
to the original C4.5 implementation improving its performance mainly for the calculation of the
entropy of continuous attributes. In this respect, we believe that \yadt is a quite paradigmatic
example of sequential, already existing, complex code of scientific and commercial interest. In
addition, \yadt is an example of extreme algorithmic sequential optimisation, which makes it
unpractical to design further optimisations. Nevertheless, the potential for improvements is vast,
and it resides in the idle core CPUs on the user's machine.

Our approach for parallelising \yadt is based on the \ff  programming framework
%\cite{fastflow-web}
\cite{fastflow:pdp:10}, a recent proposal for parallel programming over multicore platforms that
provides a variety of facilities for writing efficient lock-free parallel patterns, including
pipeline parallelism, task parallelism and Divide\&Conquer (D\&C) computations. Besides technical
features, \ff offers an important methodological approach that will lead us to parallelise \yadt
with minimal changes to the original sequential code, yet achieving up to $7\times$ boost in
performance on a Intel dual-quad core. MIPS, FLOPS  and speedup have
not to be the only metrics in software development. Human
productivity, total cost and time to solution are equally, if not
more, important.
%\cite{blog:acm:reed:2009}.

The rest of the paper is organised as follows. In Sect.~\ref{sec:fastflow}, the \ff programming
environment is introduced. We recall in Sect.~\ref{sec:yadt} the C4.5 decision tree construction
algorithm, including the main optimisations that lead to \yadt.  Then the parallelisation of \yadt
is presented in detail in Sect.~\ref{sec:parallel}, followed by experimental evaluation and
discussion in Sect.~\ref{sec:exp}. Finally, we report related works in Sect.~\ref{sec:experiences},
and summarise the contribution of the paper in the conclusions.

\section{The \ff Parallel Programming Environment}
\label{sec:fastflow}

\begin{figure}[t]
\begin{center}
\includegraphics[width=1\linewidth]{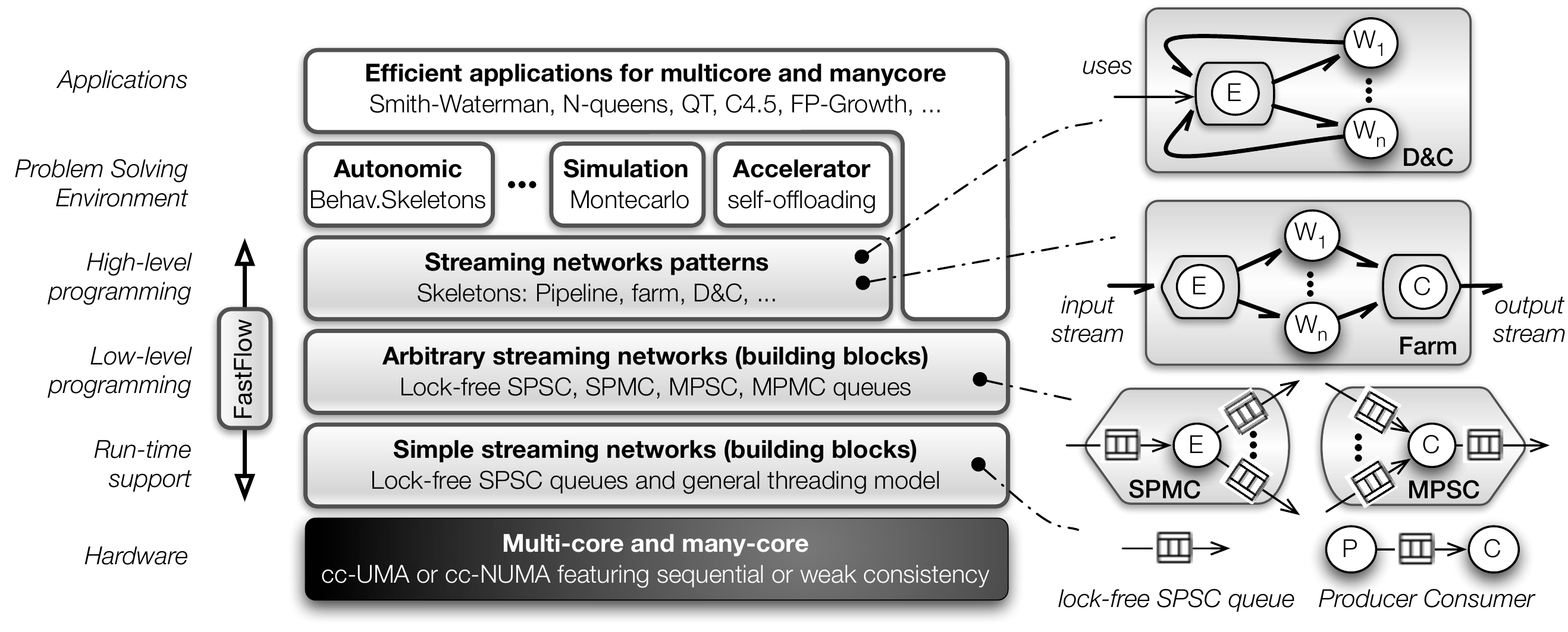}
\caption{FastFlow layered architecture with pattern examples.\label{fig:ff:architecture}}
\end{center}
\vspace{-0.5cm}
\end{figure}

FastFlow is a parallel programming framework aiming to \emph{simplify}
the development of \emph{efficient} applications for multicore
platforms, being these applications either brand new or ports of existing
legacy codes. The key vision underneath \ff is that effortless
development and efficiency  can be both achieved by raising the level
of abstraction application design, thus providing designers with a
suitable set of parallel programming patterns that can be compiled
onto efficient networks of parallel activities on the target
platforms.  To fill the abstraction gap, as shown in Fig.~\ref{fig:ff:architecture}, FastFlow is
conceptually designed as a stack of layers that progressively abstract
the shared memory parallelism at the level of cores up to the definition of useful
programming constructs and patterns.

At the lowest tier of the FastFlow system we have the
architectures that it targets: cache-coherent multiprocessors, and in
particular commodity homogeneous multicore (e.g. Intel core, AMD K10, etc.).

%Taking inspiration from Fastforward queues \cite{fastforward:ppopp:08},
%and Lamport's wait-free protocols \cite{Lamport}
The second tier provides mechanisms to define simple streaming
networks whose \emph{run-time support} is implemented through correct
and efficient lock-free Single-Producer-Single-Con\-sum\-er (SPSC)
queues. This kind of queues do not requires any lock or memory
barrier,\footnote{for Total Store Order processors, such as Intel core,
AMD 10.}
%see  \cite{fastflow-web}.}
and thus they constitute a solid ground for a low-latency synchronisation mechanism for
multicore. These synchronisations,
which  are asynchronous and non-blocking,  do not induce any additional cache invalidation as it
happens in mutual exclusion primitives (e.g. locks and interlocked operations), and thus do not add
any extra overhead.

The third tier generalises one-to-one  to one-to-many (SPMC),
many-to-one (MPSC), and many-to-many (MPMC) synchronisations and data
flows, which  are implemented using only SPSC queues and arbiter
threads. This abstraction is designed in such a way that arbitrary
networks of activities can be expressed while  maintaining
the high efficiency  of synchronisations.

The next layer up, i.e.,~\emph{high-level programming},  provides a programming framework based on
parallelism exploitation patterns (a.k.a. \emph{skeletons} \cite{cole:manifesto:02}). They are
usually categorised in three main classes: Task, Data, and Stream Parallelism. \ff specifically
focuses on Stream Parallelism, and in particular provides: \emph{farm}, \emph{farm-with-feedback}
(i.e. Divide\&Conquer), \emph{pipeline}, and their arbitrary nesting and composition. These
high-level skeletons are actually factories for parametric patterns of concurrent activities, which
can be instantiated with sequential code (within white circles in Fig.~\ref{fig:ff:architecture})
or other skeletons, then cross-optimised and compiled together with lower \ff tiers. The skeleton
disciplines concurrency exploitation within the generated parallel code: the programmer is not
required to explicitly interweave the business code with concurrency related primitives.

We refer to \cite{fastflow:pdp:10} for implementation details.
%,fastflow-web
\ff is open source available  at \url{http://sourceforge.net/projects/mc-fastflow/} under LGPLv3 license.

\section{Decision Trees: From C4.5 to \yadt} \label{sec:yadt}

A decision tree is a classifier induced by supervised learning from a relation ${\cal T}$ called
the {\em training set}. Tuples in ${\cal T}$ are called {\em cases}. An attribute $C$ of the
relation is called the {\em class}, while the remaining ones $A_1, \ldots, A_m$ are called the {\em
predictive attributes}. The domain of an attribute $dom(A_i)$ can be discrete, namely a finite set
of values, or continuous, namely the set of real numbers. Also, the special value {\em unknown} is
allowed in $dom(A_i)$ to denote unspecified or unknown values. The domain of the class $dom(C) = \{
c_1, \ldots, c_{NC}\}$ is discrete and it does not include the unknown value.

A {\em decision tree} is a tree data structure consisting of {\em decision nodes} and {\em leaves}.
A leaf specifies a class value.  A decision node specifies a {\em test} over one of the predictive
attributes, which is called the attribute {\em selected} at the node. For each possible outcome of
the test, a child node is present. A test on a discrete attribute $A$ has $h$ possible outcomes $A
= d_1$, \ldots, $A = d_h$, where $d_1, \ldots d_h$ are the known values in $dom(A)$. A test on a
continuous attribute has 2 possible outcomes, $A \leq t$ and $A > t$, where $t$ is a {\em
threshold} value determined at the node.

\subsection{The C4.5 Tree-Induction Algorithm}

The C4.5 decision tree induction algorithm \cite{Qui93} is a constant reference in the development
and analysis of novel proposals of classification models \cite{TSWY99}. The core\footnote{In this
paper, we concentrate on the {\em growth} phase of the algorithm. The subsequent {\em prune} phase
is computationally less expensive.} algorithm constructs the decision tree top-down. Each node is
{\em associated} with a set of weighted cases, where weights are used to take into account unknown
attribute values. At the beginning, only the root is present, with associated the whole training
set ${\cal T}$.  At each node a D\&C algorithm is adopted to select an attribute for splitting. We
refer the reader to the method \texttt{node::split} in Fig.~\ref{alg:node:split} from the \yadt
implementation of the algorithm.

Let $T$ be the set of cases associated at the node. For every $c \in dom(C)$, the weighted
frequency $freq(c, T)$ of cases in $T$ whose class is $c$ is computed
(\lines{\ref{alg:node:split}}{2} -- throughout the paper, we use the \textsc{\small \S M.n}
 to reference line \textsc{\small n} from the pseudo-code in
Fig. \textsc{\small M}). If
all cases in $T$ belong to the same class %$c_j$
or the number of cases in $T$ is less than a certain value then the node is set to a leaf
(\linesint{\ref{alg:node:split}}{3}{4}). If $T$ contains cases belonging to two or more classes,
then the {\em information gain} of each attribute at the node is calculated
(\linesint{\ref{alg:node:split}}{6}{7}). Since the information gain of a discrete attribute
selected in an ancestor node is necessarily $0$, the number of attributes to be considered at a
node is variable (denoted by \texttt{getNoAtts} in \lines{\ref{alg:node:split}}{6}).

For a discrete attribute $A$, the information gain of splitting $T$ into
subsets $T_1, \ldots, T_h$, one for each known value of $A$, is calculated
\footnote{as follows: $gain(T, T_1, \ldots, T_h) = info(T) - \sum_{i=1}^h
\frac{|T_i|}{|T|}\times info(T_i)$, where $info(S) = - \sum_{j=1}^{NC}
\frac{freq(c_j, S)}{|S|}\times log_2(\frac{freq(c_j, S)}{|S|})$ is the entropy
function.}. For $A$ continuous, cases in $T$ with known value for $A$ are first
ordered w.r.t.~such an attribute. Let $v_1, \ldots, v_k$ be the ordered values
of $A$ for cases in $T$. Consider for $i \in [1, k-1]$ the value $v = (v_i +
v_{i+1})/2$ and the splitting of $T$ into cases $T_1^v$ whose value for the
attribute $A$ is lower or equal than $v$, and cases $T_2^v$ whose value is
greater than $v$. For each value $v$, the information gain $gain_v$ is computed
by considering the splitting above. The value $v'$ for which $gain_{v'}$ is
maximum is set to be the {\em local threshold} and the information gain for the
attribute $A$ is defined as $gain_{v'}$.

The attribute $A$ with the highest information gain is selected for the test at the node
(\lines{\ref{alg:node:split}}{8}). When $A$ is continuous, the {\em threshold} of the split is
computed (\linesint{\ref{alg:node:split}}{9}{10}) as the greatest value of $A$ in the {\em whole}
training set ${\cal T}$ that is below the local threshold. Finally, let us consider the generation
of the child nodes (\linesint{\ref{alg:node:split}}{12}{14}). When the selected attribute $A$ is
discrete, a child node for each known value from $dom(A)$ is created, and cases in $T$ are
partitioned over the child nodes on the basis of the value of attribute $A$. When $A$ is continuous
two child nodes are created, and cases from $T$ with known value of $A$ are partitioned accordingly
to the boolean result of the test $A \leq t$, where $t$ is the threshold of the split. Cases in $T$
whose value for attribute $A$ is unknown are added to the set of cases of every child, but their
weights are rebalanced.
\renewcommand*\thelstnumber{\ref{alg:node:split}.\oldstylenums{\the\value{lstnumber}}}
\begin{figure}[t]
\begin{minipage}[t]{0.5\linewidth}
\begin{Bench}{}{}
void node::split() {
 computeFrequencies();
 if (onlyOneClass() || fewCases())
  set_as_leaf();
 else {
  for(int i=0;i<getNoAtts();++i)
   gain[i]= gainCalculation(i);
  int best = argmax(gain);
  if (attr[best].isContinuous())
   findThreshold(best);
  ns=attr[best].nSplits();
  for(int i=0;i<ns;++i)
   childs.push_back(
      new node(selectCases(best,i)));
 }
}
\end{Bench}
\vspace{1.15cm} \caption{The original \yadt node splitting procedure.\label{alg:node:split}}
\end{minipage}
\hspace{0.5ex}
\renewcommand*\thelstnumber{\ref{alg:node:splitff}.\oldstylenums{\the\value{lstnumber}}}
\begin{minipage}[t]{0.5\linewidth}
%\lstset{firstnumber=18}
\begin{Bench}{}{}
bool node::splitPre() {
 computeFrequencies();
 if (onlyOneClass() || fewCases()) {
  set_as_leaf();
  return true;
 }
 return false;
}
void node::splitAtt(i) {
 gain[i]= gainCalculation(i);
}
void node::splitPost() {
 int best = argmax(gain);
 if (attr[best].isContinuous())
  findThreshold(best);
 ns=attr[best].nSplits();
 for(int i=0;i<ns;++i)
  childs.push_back(
     new node(selectCases(best,i)));
}
\end{Bench}
\caption{Partitioning of the
  \texttt{node::split} method into three steps.\label{alg:node:splitff}}
\end{minipage}
\vspace{-0.4cm}
\end{figure}

\subsection{From C4.5 to \yadt}

The original Quinlan's implementation of C4.5 maintains the training set as an array of cases. Each
case is an array of attribute values. The decision tree is grown depth-first. The computation of
information gain takes $O(r)$ operations for discrete attributes, where $r = |T|$ is the number of
cases at the node;  and $O(r\,log\,r)$ operations for continuous attributes, where sorting is the
predominant task. Finally, searching for the threshold of the selected continuous attribute
(\lines{\ref{alg:node:split}}{10}) requires $O(|{\cal T}|)$ operations, where $|{\cal T}|$ is the
number of cases in the whole training set. This linear search prevents the implementation being
truly a D\&C computation.

Efficient C4.5 (EC4.5) \cite{ec45} is a patch software improving the efficiency of C4.5 in a number
of ways. Continuous attribute values in a case are stored as indexes to the pre-sorted elements of
the attribute domain. This allows for adopting a binary search of the threshold in the set of
domain values at \lines{\ref{alg:node:split}}{10}, with a computational cost of $O(log\,d)$
operations where $d = max_i |dom(A_i)|$. At each node, EC4.5 calculates the information gain of
continuous attributes by choosing the best among three strategies accordingly to an analytic
comparison of their efficiency: the first strategy adopts {\em quicksort}; the second one adopts
{\em counting sort}, which exploits the fact that in lower nodes of the tree continuous attributes
ranges tend to be narrow; the third strategy calculates the local threshold using a main-memory
version of the RainForest \cite{GRG00} algorithm, without any sorting.

\yadt \cite{yadt} is a from scratch C++ implementation of C4.5. It inherits the optimisations of
EC4.5, and adds further ones, e.g.,~searching the local threshold for continuous attributes by
considering splittings at {\em boundary} values (Fayyad and Irani method). Concerning data
structures, the training set is now stored by columns, not by rows, since most of the computations
scan data by attribute values. Most importantly, the object oriented design of \yadt allows for
encapsulating the basic operations on nodes into a C++ class, with the advantage that the growing
strategy  of the decision tree can now be a parameter (depth first, breadth first, or any other
top-down growth). By default, \yadt adopts a breadth first growth -- which has a less demanding
main memory occupation. Its pseudo-code is shown in Fig.~\ref{alg:tree:build} as method
\texttt{tree::build}. Experiments from \cite{ec45,yadt} show that \yadt reaches up to $10\times$
improvement over C4.5 with only 1/3 of its memory occupation.

\renewcommand*\thelstnumber{\ref{alg:tree:build}.\oldstylenums{\the\value{lstnumber}}}
\begin{figure}[t]
\begin{minipage}[t]{0.48\linewidth}
\begin{Bench}{}{}
void tree::build() {
 queue<node *> q;
 node *root = new node( allCases );
 q.push(root);
 while( !q.empty() ) {
  node *n = q.front();
  q.pop();
  n->split();
  for(int i=0;i<n->nChilds();++i)
   q.push( n->getChild(i) );
 }
}

\end{Bench}
\caption{\yadt tree growing procedure.\label{alg:tree:build}}
\end{minipage}
\hspace{0.5ex}
\renewcommand*\thelstnumber{\ref{alg:tree:buildff}.\oldstylenums{\the\value{lstnumber}}}
\begin{minipage}[t]{0.48\linewidth}
\begin{Bench}{}{}
void tree::build_ff() {
 node *root = new node( allCases );
 E=new ff_emitter(root,PAR_DEGREE);
 std::vector<ff_worker*> w;
 for(int i=0;i<PAR_DEGREE;++i)
  w.push_back( new ff_worker());
 ff_farm<ws_scheduler>
    farm(PAR_DEGREE*QSIZE);
 farm.add_workers(w);
 farm.add_emitter(E);
 farm.wrap_around();
 farm.run_and_wait_end();
}
\end{Bench}
\caption{\yadtff D\&C setup.\label{alg:tree:buildff}}
\end{minipage}
\end{figure}

\renewcommand*\thelstnumber{\ref{alg:build_node}.\oldstylenums{\the\value{lstnumber}}}
\begin{figure}[t]
\begin{Bench2}{}{}
void * ff_emitter::svc(void * task) {
 if (task == NULL) {
  task=new ff_task(root,BUILD_NODE);
  int r = root->getNoCases();
  setWeight(task, r);
  return task;
 }
 node *n = task->getNode();
 nChilds = n->nChilds();
 if (noMoreTasks() && !nChilds)
  return NULL;
 for(int i=0; i < nChilds; i++) {
  node *child = n->getChild(i);
  ctask=new ff_task(child,BUILD_NODE);
  int r = child->getNoCases();
  setWeight(ctask, r);
  ff_send_out(ctask);
 }
 return FF_GO_ON;
}

void * ff_worker::svc(void * task) {
 node *n = task->getNode();
 n->split();
 return task;
}
\end{Bench2}
\vspace{10pt} \caption{Emitter and Worker definition for the NP strategy.} \label{alg:build_node}
\vspace{-0.2cm}
\end{figure}

\renewcommand*\thelstnumber{\ref{alg:build_att}.\oldstylenums{\the\value{lstnumber}}}
\begin{figure}
\begin{Bench2}{}{}
void * ff_emitter::svc(void * task) {
 if (task == NULL ) {
  if (root->splitPre()) return NULL;
  int r = root->getNoCases();
  int c = root->getNoAtts();
  for(int i=0;i<c;++i) {
   task=new ff_task(root,BUILD_ATT);
   task->att = i;
   setWeight(task, r);
   ff_send_out(task);
  }
  root->attTasks = c;
  return FF_GO_ON;
 }
 node *n = task->getNode();
 if (task->isBuildAtt()) {
  if (--n->attTasks>0)
   return FF_GO_ON;
  n->splitPost();
 }
 nChilds = n->Childs();
 if (noMoreTasks() && !nChilds)
  return NULL;
 for(int i=0; i < nChilds; i++) {
  node *child = n->getChild(i);
  int r = child->getNoCases();
  int c = child->getNoAtts();
  if (!buildAttTest(r,c)) {
   ctask=new ff_task(child,BUILD_NODE);
   setWeight(ctask, r);
   ff_send_out(ctask);
  } else {
   if (child->splitPre()) continue;
   for(int j=0;j<c;++j) {
    ctask=new ff_task(child,BUILD_ATT);
    ctask->att = j;
    setWeight(ctask, r);
    ff_send_out(ctask);
  }
  child->attTasks = c;
 }
 return FF_GO_ON;
}

void * ff_worker::svc(void * task) {
 node *n = task->getNode();
 if (task->isBuildAtt())
  n->splitAtt(task->att);
 else n->split();
 return task;
}
\end{Bench2}
\vspace{10pt}
\caption{Emitter and Worker definition for the NAP strategy.}
\label{alg:build_att}
 \vspace{-0.4cm}
\end{figure}

\section{Parallelising \yadt}
\label{sec:parallel}

We propose a parallelisation of \yadt, called \yadtff, obtained by stream parallelism. Each
decision node is considered a task that generates a set of sub-tasks; these tasks are arranged in a
stream that flows across a \emph{farm-with-feedback} skeleton which implements the D\&C paradigm.
The \ff D\&C schema is shown in the top-right corner of Fig.~\ref{fig:ff:architecture}. Tasks in
the stream are scheduled by an \emph{emitter} thread towards a number of \emph{worker} threads,
which process them in parallel and independently, and return the resulting tasks back to the
emitter. For the parallelisation of \yadt, we adopt a two-phases strategy: first, we accelerate the
\texttt{tree::build} method (see Fig.~\ref{alg:tree:build}) by exploiting task parallelism among
node processing, and we call this strategy {\em Nodes Parallelisation} (NP); then, we add the
parallelisation of the \texttt{node::split} method (see Fig. \ref{alg:node:split}) by exploiting
parallelism also among attributes processing, and we call such a strategy {\em Nodes \& Attributes
Parallelisation} (NAP). The two strategies share the same basic setup method,
\texttt{tree::build\_ff} shown in Fig.~\ref{alg:tree:buildff}, which creates an emitter object
(\linesint{\ref{alg:tree:buildff}}{2}{3}) and an array of  worker objects
(\linesint{\ref{alg:tree:buildff}}{4}{6}). The size of the array, \texttt{PAR\_DEGREE}, is the
parallelism degree of the farm. The root node of the decision tree is passed to the constructor of
the emitter object, so that the stream can be initiated from it. The overall farm parallelisation
is managed by the \ff layer through a \texttt{ff\_farm} object, which creates feedback channels
between the emitter and the workers (\linesint{\ref{alg:tree:buildff}}{7}{11}). Parameters of
\texttt{ff\_farm} include: the size \texttt{QSIZE} of each worker input queue, and the scheduling
policy (\texttt{ws\_scheduler}), which is based on tasks weights. Basically, such a policy assigns
a new task to the worker with the lowest total weight of tasks in its own input FIFO queue. The
emitter class \texttt{ff\_emitter} and the worker class \texttt{ff\_worker} define the behaviour of
the farm parallelisation through the class method \texttt{svc} (short name for {\em service}) that
is called by the \ff run-time to process input tasks. Different parallelisation strategies can be
defined by changing only these two methods. The implementation of the NP and the NAP strategies are
shown in Fig.~\ref{alg:build_node} and Fig.~\ref{alg:build_att} respectively.

\paragraph{NP strategy (Fig.~\ref{alg:build_node}).} At start-up the \texttt{ff\_emitter::svc} method
is called by the \ff run-time with a NULL parameter (\lines{\ref{alg:build_node}}{2}). In this
case, a task for processing the root node is built, and its weight is set to the number of cases at
the root (\linesint{\ref{alg:build_node}}{3}{5}). Upon receiving in input a task coming from a
worker, the emitter checks the termination conditions (\lines{\ref{alg:build_node}}{10}), and then
produces in output the sub-tasks corresponding to the children of the node
(\linesint{\ref{alg:build_node}}{12}{18}). The \texttt{ff\_send\_out} method of the \ff runtime
allows for queueing tasks without returning from the method. Finally, the \texttt{FF\_GO\_ON} tag
in the return statement (\lines{\ref{alg:build_node}}{19}) tells the run-time that the computation
is not finished (this is stated by returning NULL), namely further tasks must be waited for from
the input channel. The \texttt{ff\_worker::svc} method for a generic worker
(\linesint{\ref{alg:build_node}}{22}{25}) merely calls the node splitting algorithm
\texttt{node::split}, and then it immediately returns the computed task back to the emitter. The
overall coding is extremely simple and intuitive -- almost a rewriting of the original
\texttt{tree::build} method. Moreover, it is quite generalisable to any top-down tree-growing
algorithm with greedy choice of the splitting at each node. The weighted scheduling policy is the
most specific part\footnote{It was not among \ff strategies, and it has been added as a result of
\yadtff.}; in particular, for the use of weights that are linear in the number of cases at the
node. This is motivated by the experimental results of \cite[Fig.~1]{ec45}, which show how the
\yadt implementation of \texttt{node::split} exhibits a low-variance elapsed time per case for the
vast majority of nodes.

\paragraph{NAP strategy (Fig.~\ref{alg:build_att}).} The NAP strategy builds over NP. For a given
decision node, the emitter follows a D\&C parallelisation over its children, as in the case of the
NP strategy. In addition, for each child node, the emitter may decide to parallelise the
calculation of the information gains in the \texttt{node::split} method
(\linesint{\ref{alg:node:split}}{6}{7}). In such a case, the stopping criterion at
\lines{\ref{alg:node:split}}{3} must be evaluated prior to the parallelisation, and the creation of
the child nodes must occur after all the information gains are computed. This leads to partitioning
the code of \texttt{node::split} into three methods, as shown in Fig.~\ref{alg:node:splitff}.

For the root node, attribute parallelisation is always the case
(\linesint{\ref{alg:build_att}}{3}{10}). A task with label \texttt{BUILD\_ATT} is constructed for
each attribute, with the field \texttt{att} recording the attribute identifier (the index
\texttt{i}). Tasks are weighted and queued. The information about how many tasks are still to be
completed is maintained in the \texttt{attTasks} field of the decision node -- such a field is
added to the original \texttt{node} class. Upon receiving in input a task coming from a worker, the
emitter checks whether it concerns the processing of an attribute
(\lines{\ref{alg:build_att}}{16}). If this is the case (\linesint{\ref{alg:build_att}}{17}{20}),
the \texttt{attTasks} counter is decremented until the last attribute task arrives, and then the
\texttt{node::splitPost} method is called to evaluate the best split. At this point, the emitter is
given a processed node (either from a worker, or as the result of the \texttt{node::splitPost}
call). Unless the termination conditions occur (\lines{\ref{alg:build_att}}{22}), the emitter
proceeds with outputing tasks. The \texttt{buildAttTest} at \lines{\ref{alg:build_att}}{28}
controls for each child node whether to generate a single node processing task, or one attribute
processing task for each attribute at the child node. In the former case
(\linesint{\ref{alg:build_att}}{29}{31}), we proceed as in the NP strategy; in the latter case
(\linesint{\ref{alg:build_att}}{33}{38}), we proceed as for the root node\footnote{Notice that
tasks for node processing are labelled with \texttt{BUILD\_NODE}, while tasks for attribute
processing are labelled with \texttt{BUILD\_ATT}}. Based on the task label, the
\texttt{ff\_worker::svc} method for a generic worker (\linesint{\ref{alg:build_att}}{46}{50})
merely calls the node splitting procedure or the information gain calculation for the involved
attribute.

Let us discuss in detail two relevant issues. Let $r$ be the number of cases and $c$ the number of
attributes at the node.

The first issue concerns task weights. Node processing tasks are weighted with $r$
(\lines{\ref{alg:build_att}}{30}), as in the NP strategy.  Although attribute processing tasks have
a finer grain, which suggests a lower weight, there exists a synchronisation point -- all attribute
tasks must be processed before the emitter can generate tasks for the child nodes. By giving a
lower weight, we run the risk that all attribute tasks are assigned to the most unloaded worker,
thus obtaining a sequential execution of the attribute tasks! For these reasons, attribute
processing tasks are weighted with $r$ as well (\linespair{\ref{alg:build_att}}{9}{37}).

The second issue concerns the test \texttt{buildAttTest}, which decides whether to perform nodes or
attributes parallelisation. We have designed and experimented three cost models. Attribute
parallelisation is chosen respectively when:
\begin{itemize}
    \item $(\alpha < r)$ the number of cases is above some hand-tuned threshold value $\alpha$;

    \item $({|\cal T|} < c\,r\,log\,r)$ the average grain of node processing (quicksort is $r\,log\,r$ on
average) is higher than a threshold that is dependent on the training set. Intuitively, the
threshold should be such that the test is satisfied at the root node, which is the coarser-grained
task, and for nodes whose size is similar. Since the average grain of processing an attribute at
the root is ${|\cal T|}\,log\,{|\cal T|}$, we fix the threshold to a lower bound for such a value,
namely to ${|\cal T|}$;

    \item $({|\cal T|} < c\,r^2)$ the worst-case grain of node processing (quicksort is $r^2$) is higher than
a threshold that is dependent on the training set. As in the previous case, the threshold is set to
${|\cal T|}$. The higher value $c r^2$, however, leads to selecting attributes processing more
often than the previous case, with the result of task over-provisioning.
\end{itemize}
All tests are monotonic in the number $r$ of cases at the node. Hence, if the nodes parallelisation
is chosen for a node, then it will be chosen for all of its descendants. As we will see in Sec.
\ref{sec:exp}, the third cost model shows the best performance.

\section{Performance Evaluation}
\label{sec:exp}

\begin{table}[tb]
\begin{center}
\textsf{\scriptsize
\begin{tabular}{l@{\hspace{2ex}}r@{\hspace{2ex}}r@{\hspace{4ex}}rrr@{\hspace{4ex}}rr}
\toprule
% \multicolumn{1}{c}{\mbox{}} & \multicolumn{3}{c}{\mbox{}}  &
&&& \multicolumn{3}{c}{\centering No. of attributes} &
 \multicolumn{2}{c}{\centering Tree} \\
\cmidrule(r){4-6}
\cmidrule(l){7-8}
${\cal T}$ name& {$|{\cal T}|$} & $NC$ &
discr. & contin.& total& size &\ depth\\
\midrule
{\em Census PUMS} & 299,285 & 2 & 33 & 7 & 40  & \ 122,306 & 31\\
{\em U.S. Census} & 2,458,285 & 5 & 67 & 0 & 67 & 125,621 & 44 \\
{\em KDD Cup 99} &  4,898,431 & 23 & 7 & 34 & 41 & 2,810 & 29\\
{\em Forest Cover} & 581,012 & 7 & 44 & 10 & 54 & 41,775 & 62\\
{\em SyD10M9A} & \ 10,000,000 & 2 & 3 & 6 & 9 & 169,108 & 22\\ %\hline
\bottomrule
\end{tabular}
}
\end{center}
\caption{Training sets used in experiments, and size of the induced decision tree.}
\label{tab:spec}
 \vspace{-0.6cm}
\end{table}

In this section we show the performances obtained by \yadtff.
%the acceleration of the \yadt algorithm using FastFlow.
The datasets used in the tests with their characteristics are reported in Table~\ref{tab:spec}.
They are publicly available from the UCI KDD archive, apart from {\em  SyD10M9A} which is
synthetically generated using function 5 of the QUEST data generator. All presented experimental
results are taken performing 5 runs, excluding the higher and the lower value obtained and
computing the average of the remaining ones.

\paragraph{Experimental framework.} All experiments %reported in the following sections
were executed on two different Intel workstation architectures:
\emph{Nehalem)} a dual quad-core Xeon E5520 Nehalem (16 HyperThreads) @2.26GHz with
8MB L3 cache and 24 GBytes of main memory  with Linux x86\_64.
%CentOS 5.4.
\emph{Harpertown)} a dual quad-core Xeon E5420 Harpertown @2.5GHz 6MB L2 cache
and 8 GBytes of main memory,  with Linux x86\_64.
% CentOS 5.2.
They are a quite standard representative of current and immediately
preceding generation of (low-cost) server boxes.
%The Nehalem-based machine uses the new Intel Quickpath interconnect
%and exploits an extended version of MESI cache coherence protocol.
The Nehalem-based machine exploits Simultaneous MultiThreading (SMT, a.k.a. HyperThreading) with 2
contexts per core and the novel Quickpath interconnect equipped with  a distributed cache coherency
protocol. SMT technology makes a single physical processor appear as two logical processors for the
operating system, but all execution resources are shared between the two contexts: caches of all
levels, execution units, etc.
%instruction fetch, etc.

% with superscaler technology.
% The Harpertown machine is a previous generation multicore model. It does
% not exploit SMT and has a different cache hierarchy with respect to the
% Nehalam-based one.
% However, we have noticed that all the 5 runs exhibit
%very low variance.
%%All tested codes are available at the \ff website \cite{fastflow-web}.

\begin{figure}[t]
\begin{center}
\hspace*{-2ex}\includegraphics[width=0.52\linewidth]{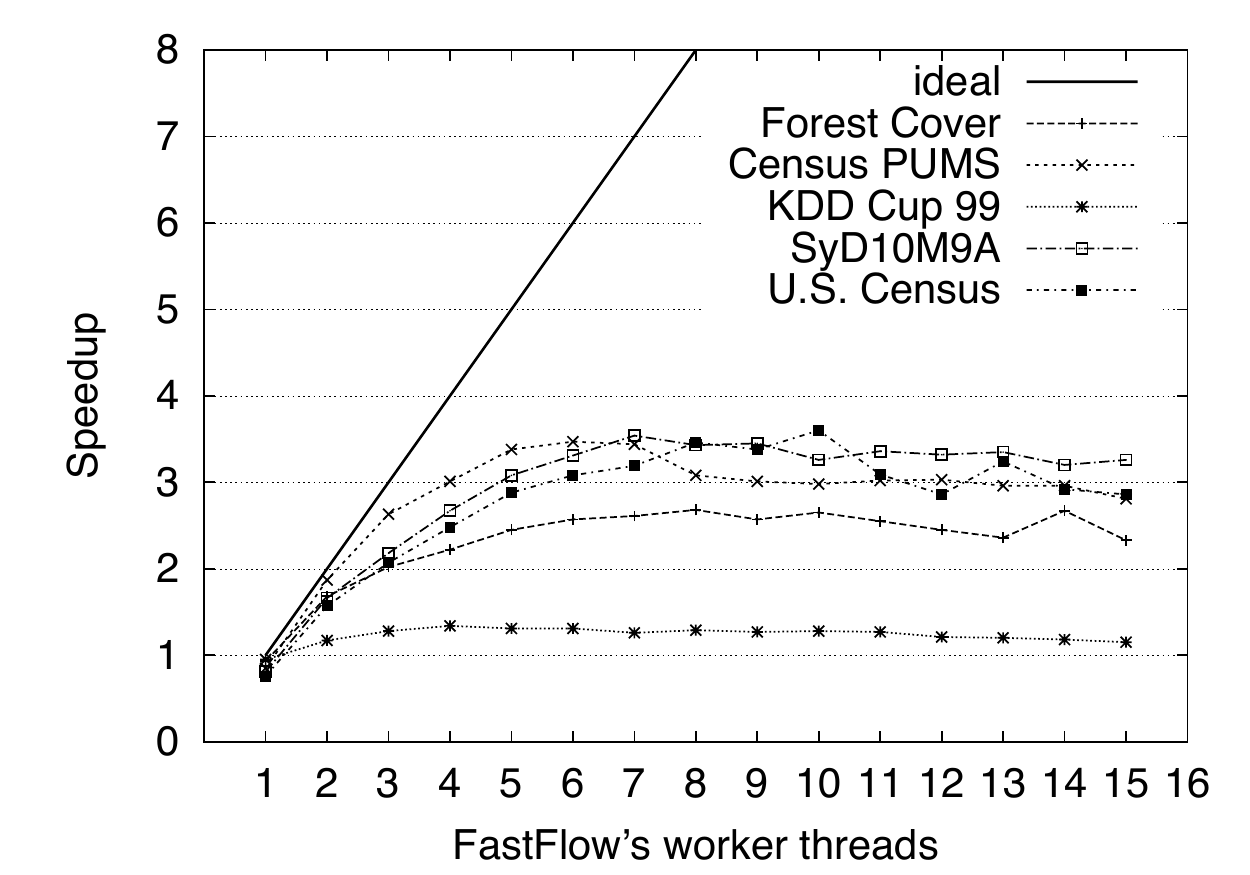}
\hspace*{-2ex}\includegraphics[width=0.52\linewidth]{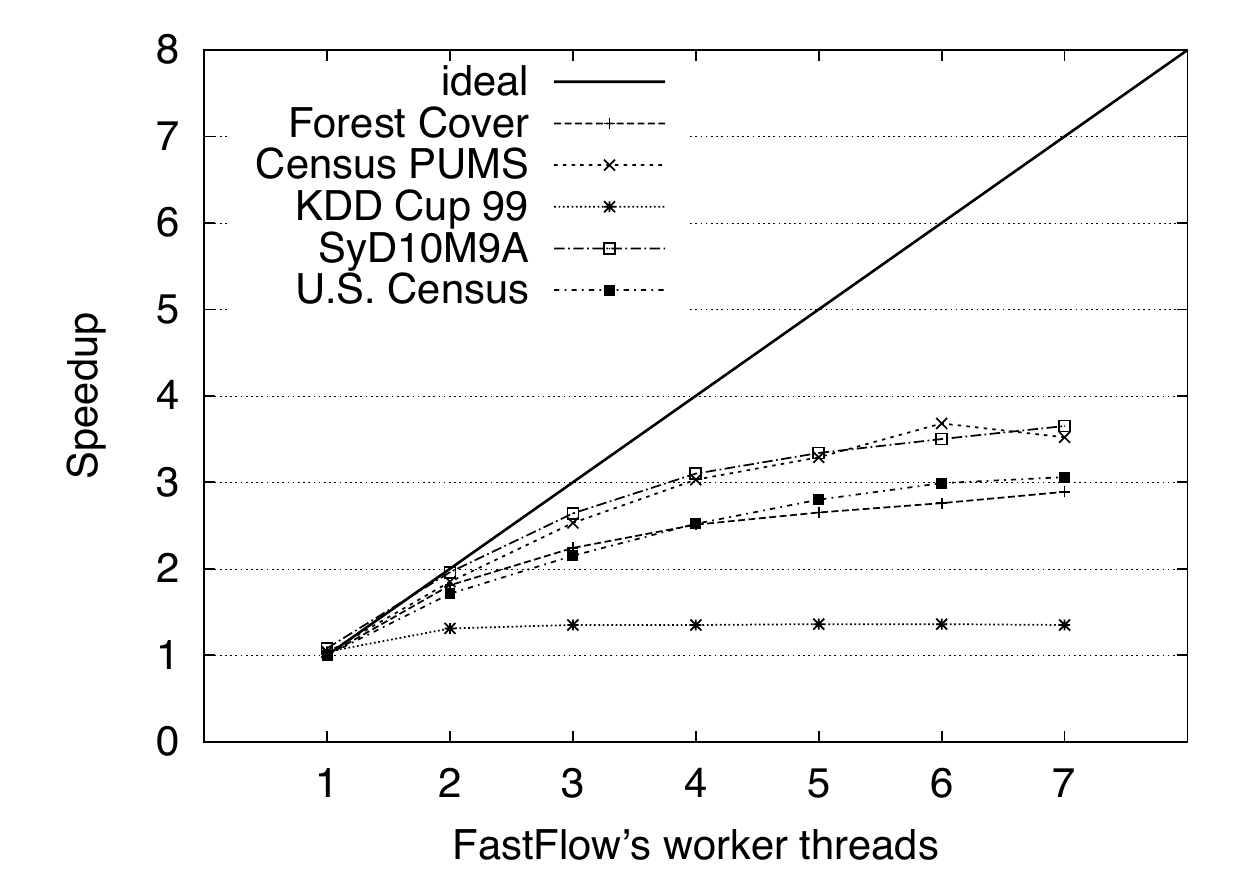}
\caption{
%\yadtff speedup when using the
NP strategy speedup. Nehalem box (left), Harpertown box (right).\label{fig:ff:onlynode}}
\end{center}
 \vspace{-0.8cm}
\end{figure}

\paragraph{Performance.} Let us start considering the {\em NP strategy}, i.e.,~the parallelisation of
nodes processing. The obtained speedup is shown in Fig.~\ref{fig:ff:onlynode}. The maximum speedup
is similar on both architectures, and quite variable from a dataset to another; it ranges from 1.34
to 3.54 (with an efficiency of $45\%$). As one would expect, exploiting inter-nodes parallelism
alone is not enough to reach a close to optimal speedup, because a large fraction of the computing
time is spent in the coarse-grained nodes (those in the higher levels of the tree), thus lacking
parallelism. This phenomenon has been already observed in previous work on the parallelisations of
decision tree construction over distributed memory architectures \cite{HSK96}. These systems,
however, suffer from load balancing problems, which we will handle later on,  and high costs of
communications, which in shared memory architectures do not occur. Summarising, although the NP
strategy yields a modest speedup, it is worth noting that the effort required to port the
sequential code was minimal.

The {\em NAP strategy} aims at increasing the available parallelism by exploiting concurrency also
in the computation of the information gain of attributes. This is particularly effective for nodes
with many cases and/or attributes, because it reduces the sequential fraction of the execution. As
presented in Sec. \ref{sec:parallel}, the emitter relies on a \emph{cost model} in order to decide
whether to adopt attributes parallelisation. We have tested the three cost models discussed in Sec.
\ref{sec:parallel}. Fig.~\ref{fig:split} shows that the test $|{\cal T}|<cr^2$ provides the best
performance for almost all datasets. This is justified by the fact that the test exhibits an higher
task over-provisioning if compared to the test $|{\cal T}|<cr\,log\,r$, and it is dataset-tailored
if compared to $\alpha < r$. In all of the remaining experiments, we use that model.

\begin{figure}[t]
\begin{center}
\hspace*{-5ex}
\begin{minipage}{0.48\linewidth}
\includegraphics[width=1.1\linewidth]{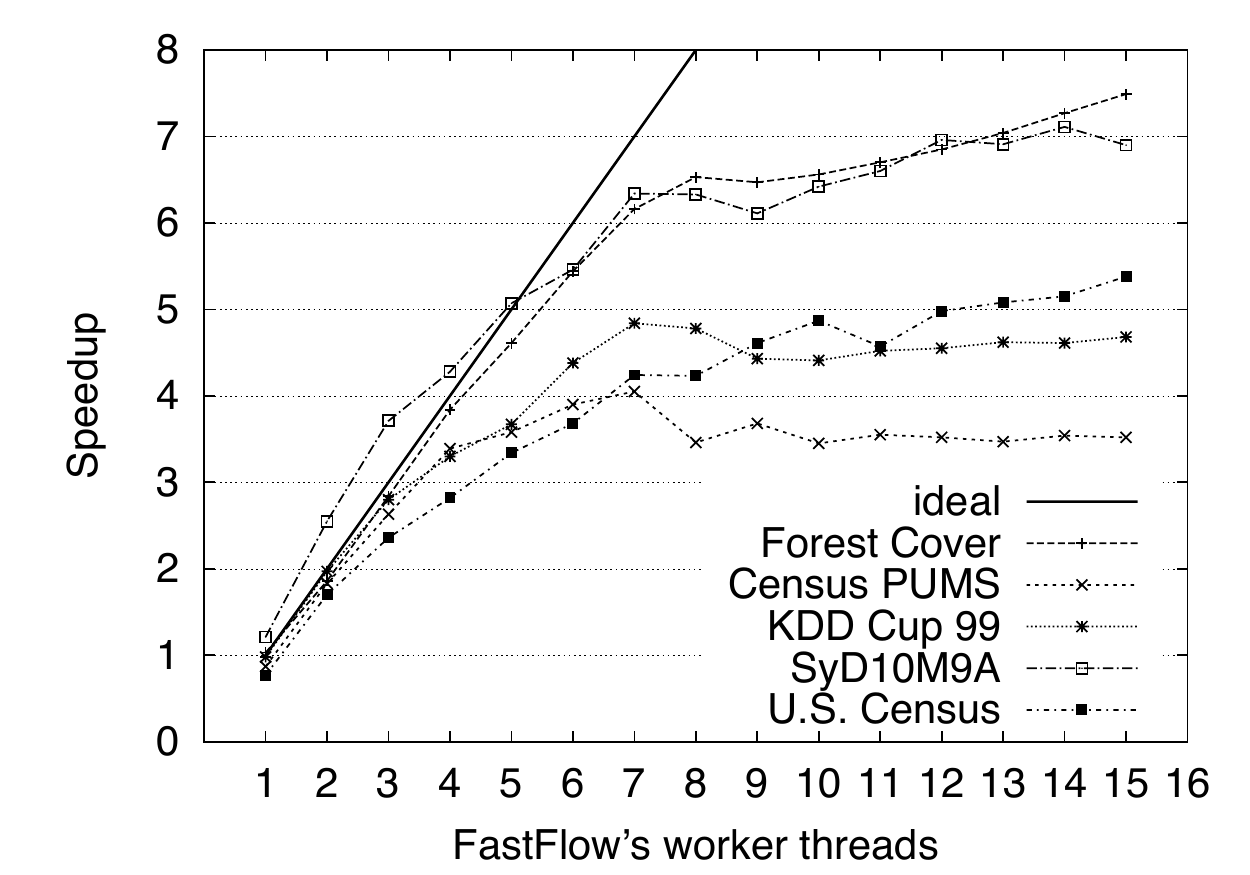}
\end{minipage}
\hspace{2ex}
\begin{minipage}{0.48\linewidth}
\includegraphics[width=1.1\linewidth]{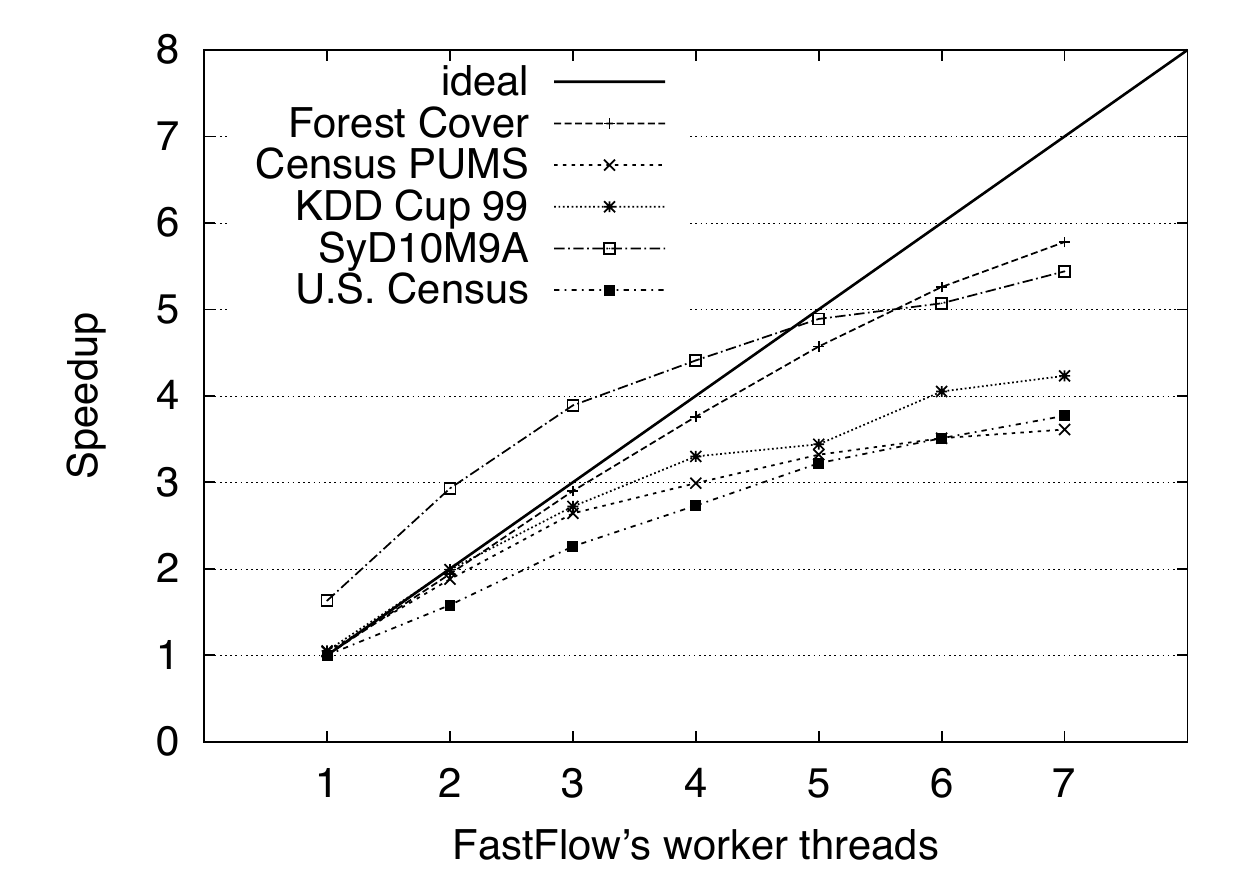}
\end{minipage}
\caption{
%\yadtff speedup when using the
NAP strategy speedup. Nehalem box (left), Harpertown box (right).\label{fig:ff:speedup}}
\end{center}
\end{figure}

The speedup of \yadtff with the NAP strategy is shown in Fig.~\ref{fig:ff:speedup}. It ranges from
4 to 7.5  (with an efficiency of $93\%$). The speedup gain over the NP strategy is remarkable. Only
for the {\em Census PUMS} dataset, the smallest dataset as for number of cases, the speedup gain is
just +12\% over NP.
% where the speedup gain over NP mainly
% depends of the number of cases of large nodes; the minimum gain
% (+12\% over NP) is exibited by  {\em Census PUMS} that has few cases (see
% Table~\ref{tab:spec}).
% The speedup obtained for the {\em
% Census PUMS} dataset is almost the same of the one obtained without
% considering the parallelism on the attributes (just 12\% higher), this
% is because the dataset has just few distinct cases thus a very small
% task computation granularity.
Notice that the \emph{SyD10M9A} dataset apparently benefits from a super-linear speedup. Actually,
this happens because the speedup is plotted against the number of farm workers. Hence, the fraction
of work done by the emitter thread is not considered, yet not negligible as shown in
Fig.~\ref{fig:breakdown}.

\begin{figure}[t]
\begin{center}
%\hspace*{-5ex}
\begin{minipage}{0.48\linewidth}
\hspace*{-2ex}\includegraphics[width=1.1\linewidth]{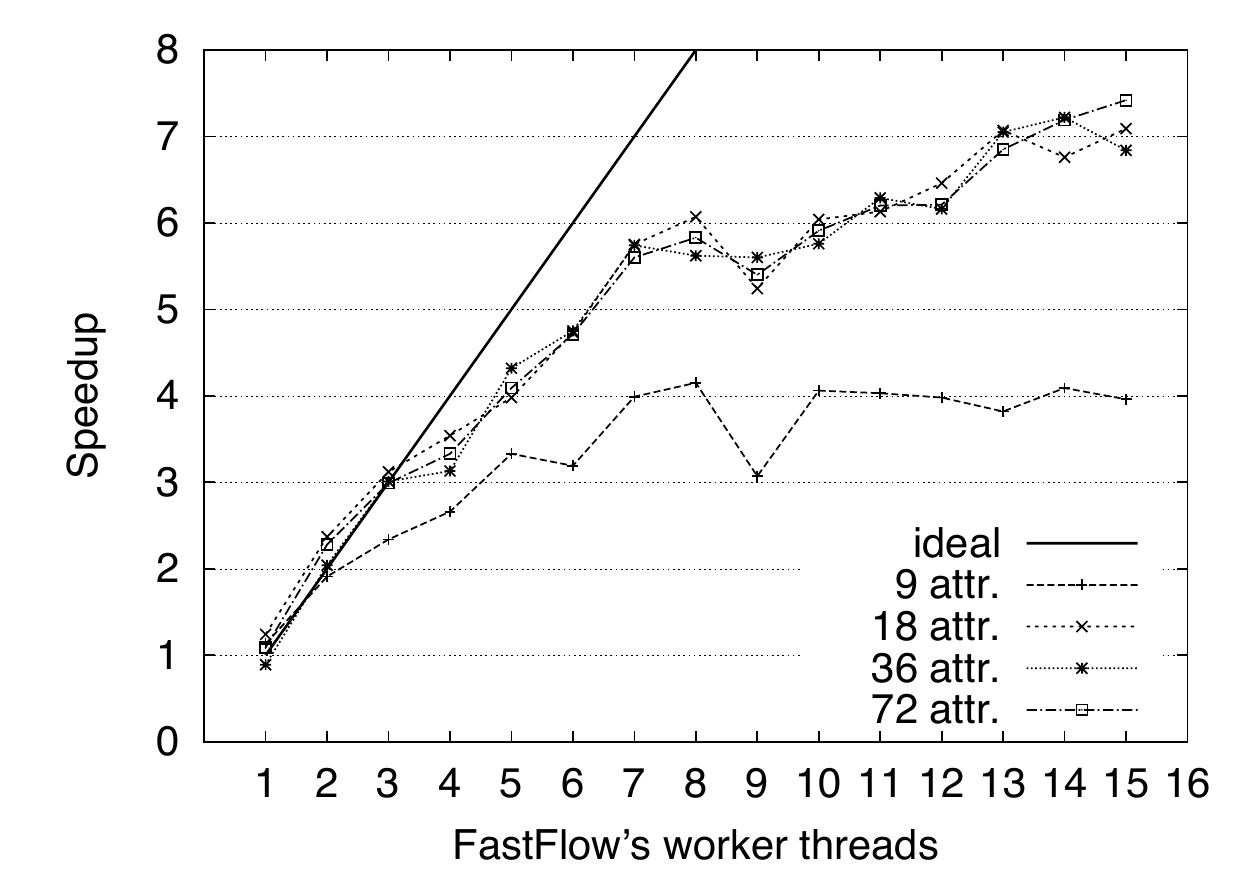} \caption{Speedup
  vs no. of attri\-bu\-tes for $1M$ sample cases from
  \emph{SyD10M9A}.\label{fig:attributes}}
\end{minipage}
\hspace{1ex}
\begin{minipage}{0.48\linewidth}
\hspace*{-2ex}\includegraphics[width=1.1\linewidth]{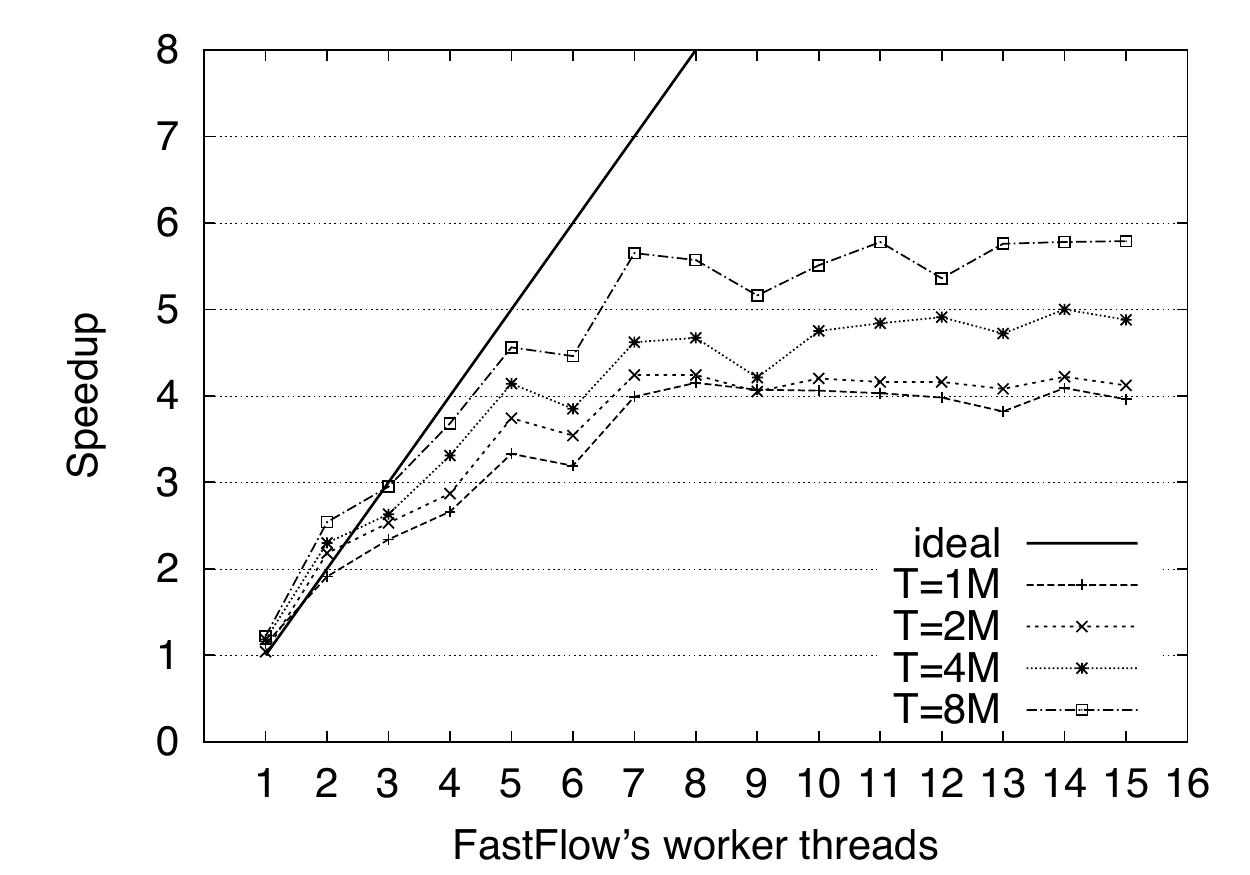} \caption{Speedup
vs no. of sample cases ($T$) from \emph{SyD10M9A}.\ \ \ \ \ \ \ \ \ \
\ \ \ \ \ \ \ \ \ \ \ \ ~\label{fig:cases}}
\end{minipage}
\end{center}
 \vspace{-0.6cm}
\end{figure}

\begin{figure}[t]
\centering
\begin{minipage}[b]{0.45\linewidth}
\textsf{\scriptsize
\begin{tabular}[t]{@{}l@{\hspace{1ex}}r@{\hspace{1ex}}r@{\hspace{2ex}}r@{}}
\toprule
&\multicolumn{3}{c}{Total Execution Time (sec.)}\\
\cmidrule{2-4}
${\cal T}$ name&$|{\cal T}|<cr^2$ &$\alpha<r$&$|{\cal T}|<cr\log r$\\
%\cmidrule(r){2-3}
%\cmidrule(r){4-5}
%\cmidrule(r){6-7}
\midrule
{\em Census PUMS}  &\textbf{0.85}&0.85& 0.91\\
{\em U.S. Census}  &\textbf{3.28}&3.51& 3.35\\
{\em KDD Cup 99}  &\textbf{3.76}&3.80& 3.77\\
{\em Forest Cover}  &\textbf{2.64}&2.66& 2.73\\
{\em SyD10M9A}  &16.90&\textbf{16.68}&18.16\\
\midrule
\end{tabular}
\begin{tabular}{@{}p{\linewidth}@{}}
Effectiveness of  {\em buildAttTest(c,r)} implementations for different attributes parallelisation
cost models.
%cost model in forecasting node size with 7 workers.
${|\cal T|}$ is the  number of cases, $c$ is the
number of node attributes, $r$ is the number of node training
cases, and $\alpha=1000$. Bold figures highlight best results.\\
\bottomrule
\end{tabular}
} \vspace*{0.5ex} \caption{Attributes parallelisation tests (Nehalem, 7 worker
threads).\label{fig:split}}
\end{minipage}
\hspace{0.4ex}
\begin{minipage}[b]{0.48\linewidth}
\hspace{-2ex}\includegraphics[width=1.1\linewidth]{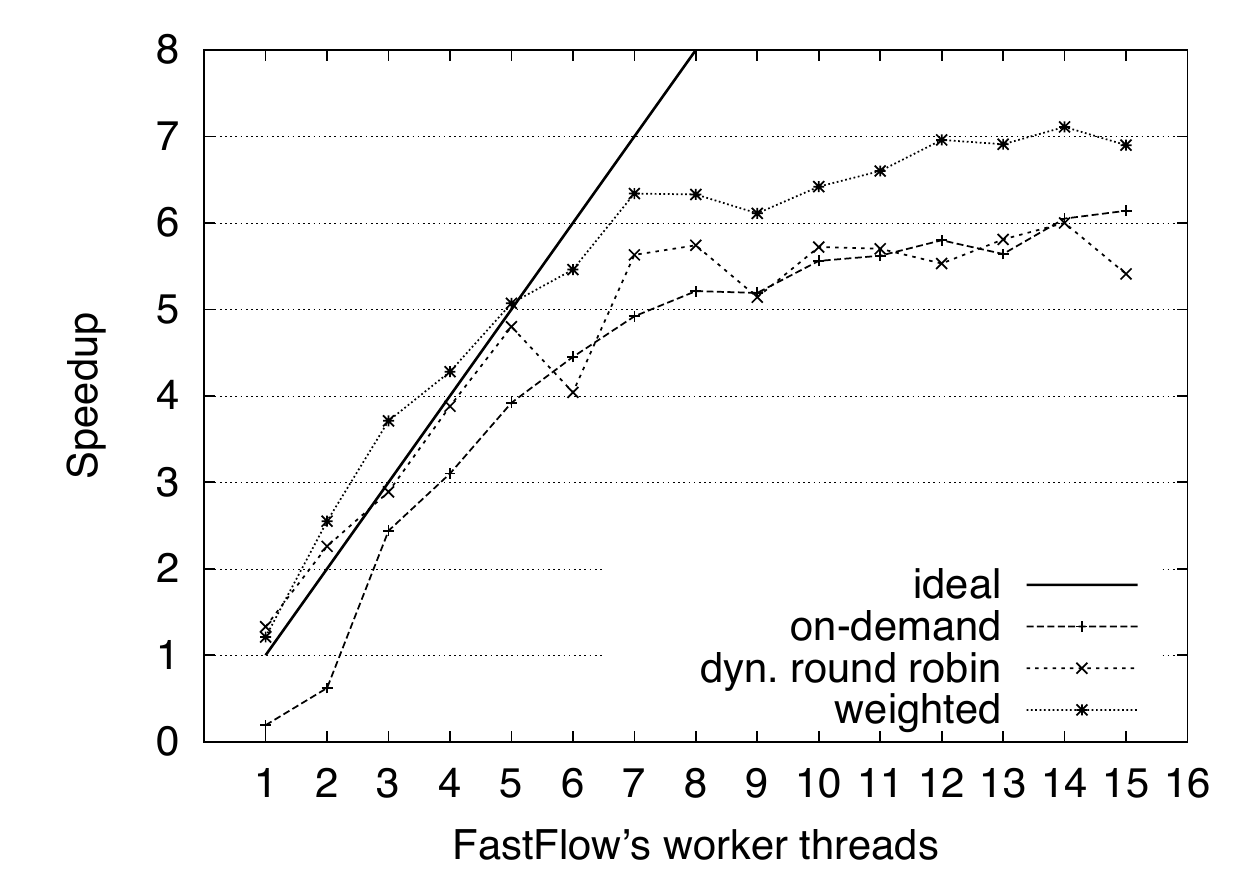}
%\vspace*{-1ex}
\caption{Speedup of different scheduling policies over \emph{SyD10M9A}.\label{fig:scheduling}}
\end{minipage}
\mbox{}\vspace{0.2cm}
\begin{minipage}[b]{0.48\linewidth}
\hspace{-2ex}\includegraphics[width=1.1\linewidth]{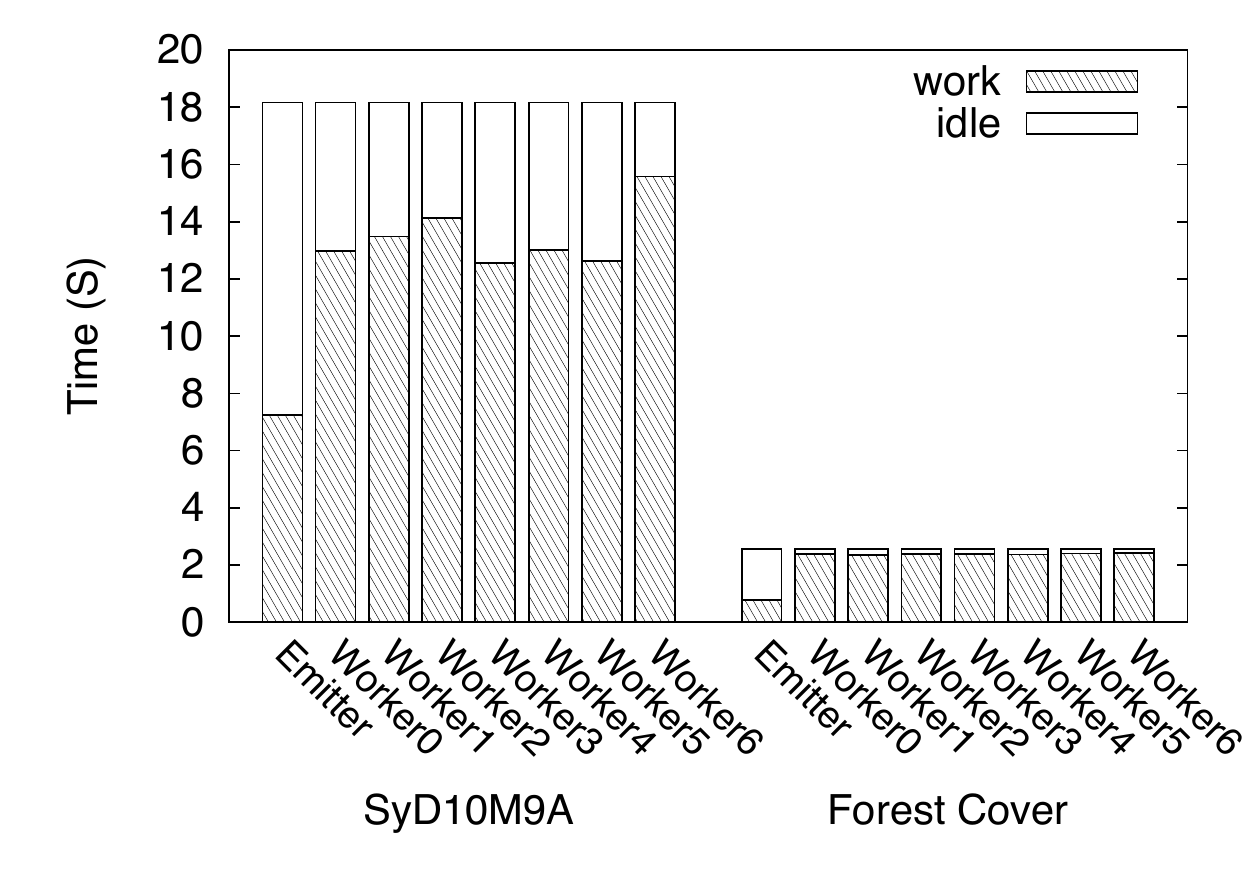}
%\vspace{-1ex}
\caption{\yadtff execution breakdown (Nehalem, 7 worker threads).\label{fig:breakdown}}
\vspace{-0.4cm}
\end{minipage}
\hspace{1.8ex}
\begin{minipage}[b]{0.48\linewidth}
\includegraphics[width=1\linewidth]{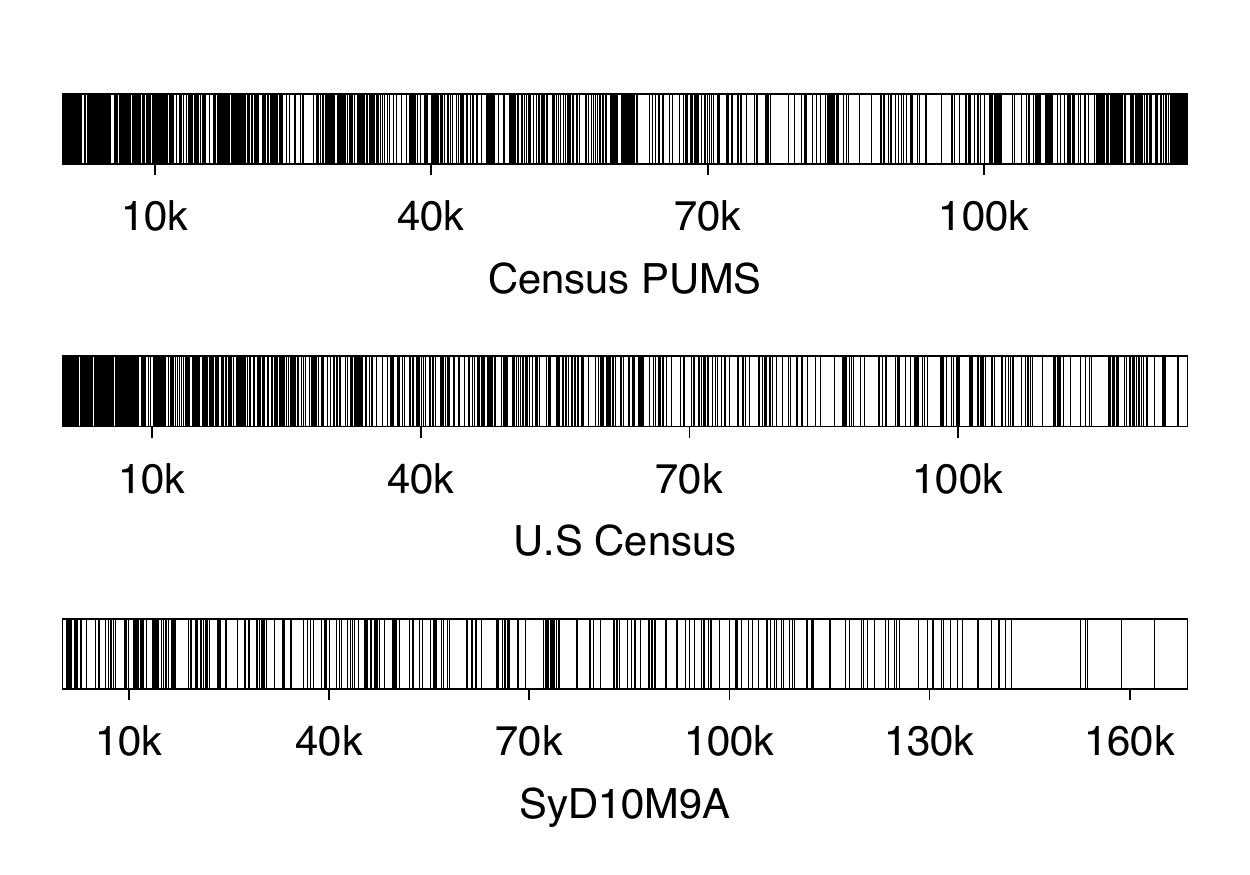}
%\vspace{-1ex}
\caption{Nodes (white) vs attributes (black) parallelisation choices.\label{fig:barcode}}
\vspace{-0.4cm}
\end{minipage}
\end{figure}

\yadtff also exhibits a good scalability with respect to both the number of attributes
(Fig.~\ref{fig:attributes}) and to the number of cases (Fig.~\ref{fig:cases}) in the training set.
The plots refer to subsets of the \emph{SyD10M9A} dataset possibly joined with randomly distributed
additional attributes. In the former case, the maximum speedup ($7\times$) is reached as soon as
the number of attributes doubles the available hardware parallelism (18 attributes for 8 cores). In
the latter case, the achieved speedup increases with the number of cases in the training set.
% The speedup variations with more than 9 attributes is not
% significant because the available HW parallelism of the underling
% architectures has been saturated.  In fact, already with 18 attributes
% we obtained a maximum speedup of 7.09 using 15 worker threads.  The
% other interesting case is the speedup obtained fixing the numbur of
% attributes and varying the number of cases. The results using 9
% attributes are shown in Fig. \ref{fig:ff:attr_coll} (right size).  The
% size of the tree with ${|\cal T|}$ = 1M is 41982 nodes with a maximum
% depth of 19, while the size of the tree with ${|\cal T|}$=8M is 147983
% nodes with a depth of 22. As expected, the bigger the dataset, the
% better speedup can be achieved.

\paragraph{Load-balancing.} The parallelisation of decision tree construction algorithms may suffer
from load balancing issues due to the difficulties in predicting the time needed for processing a
node or a sub-tree. This is exacerbated in the parallelisation of the original C4.5 implementation,
because of the linear search of the threshold (\lines{\ref{alg:node:split}}{10}).
Fig.~\ref{fig:breakdown} shows that load balancing  is not a critical issue for \yadtff with the
NAP strategy. We motivate this by two main reasons: 1) the NAP strategy produces a significant
over-provisioning of tasks with respect to the number of cores; these tasks continuously flow (in a
cycle) from the emitter to the workers and they are subject to online scheduling within the
emitter; 2) \ff communications are asynchronous and exhibit very low overhead
\cite{fastflow:pdp:10}. This makes it possible to sustain all the workers with tasks to be
processed for the entire run. This also reduces the dependence of the achieved speedup from the
effectiveness of the scheduling policy. Nevertheless, such dependence exists.

Fig.~\ref{fig:scheduling} shows results for three different scheduling policies: 1) Dynamic
Round-Robin (DRR); 2) On-Demand (OD); 3) Weighted Scheduling (WS).   The DRR policy schedules a
task to a worker in a round-robin fashion, skipping workers with full input queue (with size set to
4096).
% It is not a
% strict round-robin policy because if worker's input queue is full and
% so the task cannot be immediately scheduled to the worker, instead of
% waiting for a free room in the queue, it is chosen the next worker. In
% our tests the input queue of each farm worker has been set to 4096
% entries.
The OD policy is a fully online scheduling, i.e.,~a DDR policy where each worker has an input queue
of size $1$.
% consists in scheduling a task upon receiving a
% request from the worker. In \ff, this policy is implemented using an
% input queue of size 1 for each worker, and considering a request
% message the condition queue empty.
The WS policy is a user-defined scheduling that can be set up by assigning weights to tasks through
calls to the \texttt{setWeight} method. \yadtff adopts a WS policy, with the weight of a task set
to the number $r$ of cases at the node.
% In \yadtff we introduced the idea of WS that consists in to schedule
% each task to the worker that is considered the one under lighter
% pressure at scheduling decision time.
% The load produced by the scheduled task is forecasted by way of
% its \emph{weight}, i.e. the number of training cases associated with the
% node (${r}$).
%The expected workers' load is continually updated within the emitter by
%tracing scheduled but not completed tasks.
% The weight consists in the number of
% training cases associated with a node (${r}$).  A node is scheduled to
% the worker which has the lower load, where the load is computed as the
% sum of all the weights currently present in the worker input queue
% (olso in this case the input queue size of each worker is set to 4096
% entries).

It is immediate to observe from Fig.~\ref{fig:scheduling} that all the scheduling policies are
fairly efficient. WS exhibits superior performance because it is tailored over the \yadtff
algorithm; it actually behaves as a quite efficient online scheduling. Finally, we show in
Fig.~\ref{fig:barcode} how often nodes parallelisation has been chosen by the emitter against the
attributes parallelisation (we recall that the test $|{\cal T}|<cr^2$ was fixed). Black stripes
lines in the figure denote attributes parallelisation choices whereas white stripes denote nodes
parallelisation ones. As expected, the former occurs more often when processing the top part of the
decision tree (from left to the right, in the figure).

\captionsetup[table]{position=bottom}

\paragraph{Simultaneous MultiThreading.} We briefly evaluate the benefits achieved using the Nehalem
HyperThreaded box. SMT is essentially a memory latency hiding technique that is effective when
different threads in a core exhibit a shared working set that induces high cache hit rate. However,
even in non-ideal conditions, SMT is able to moderately increase instructions per clock-cycle
count, hence the overall performance, by partially hiding costly memory operations with threads
execution. The comparison between the two graphs in Fig.~\ref{fig:ff:speedup} shows that several
datasets benefit of (about) 30\% improvement due to SMT; some others, such as \emph{Census PUMS}
and {\em KDD Cup}, show only a modest benefit (about 12\%). These figures match the expected
benefit for this kind of architectures. As future work, we believe that the effectiveness of SMT
can be further improved by devising a \emph{cache-aware} weighted scheduling policy. This will not
affect the general structure of the code but only the calls the \texttt{setWeight} method
(\lines{\ref{alg:build_node}}{16}).

\section{Related Work}
\label{sec:experiences}

Over the last decade, parallel computing aimed at addressing three main classes of data mining
issues: 1) solve inherently distributed mining problems, e.g.,~mining of datasets that are bound to
specific sites due to privacy issues; 2) manage larger datasets by exploiting the aggregate
memories of different machines; 3) decrease the processing time of mining algorithms. In many
cases, the latter two issues have been jointly addressed by trying to bring in-core datasets that
are out-of-core on a single machine. This approach, which often requires the redesign of the
algorithms or the introduction of new scalable data structures, is
 loosing interest with the ever-increasing availability
of main memory space. Our work distinguishes from this approach, even
if it clearly belongs to the third class.
%To the best our knoweledge, no previous work investigated

% Several works focusing on parallelism has been presented over the years for
% data mining algorithms.
% Many of these works try to exploit the aggregate memories
% of distributed memory architectures to face very big out-of-core datasets. This approach,
% has often required the modification of the algorithms or the introduction of
% new scalable data structures. Conversely, other studies, have tried to
% decrease the computation time of in-core dataset without any
% significant algorithms modification.
% The work presented in this paper falls in this latter class of studies.
% Today, due to the very low costs of memories, machines with hundreds of gigabytes
% of memory are quite common and probably this is one of the motivation why there is a
% diminishing number of proposals for parallel implementations of data mining algorithms.

Considering  the parallelisation methodology, related works can be categorised as follow: 1)
exploiting attributes parallelism by partitioning the training set by columns and then adopting
data parallelism \cite{SAM96,JKK98}; 2) exploiting nodes parallelism by independently building
different nodes or sub-trees adopting task parallelism \cite{DBLP:journals/pc/CoppolaV02}; 3)
combining the two above in various fashions \cite{SAR99,ZHA99}.

%% Several different parallel strategies for decision tree classifiers have been
%% studied in the literature. Some of these strategies exploit data parallelism
%% available among the different attributes \cite{MAR96,SAM96,JKK98}.
%% The data parallel approach is efficient at building the top parts of the
%% tree where split points can be
%% evaluated in parallel.

%% Other strategies rely on task parallelism, i.e. the
%% independent building of different subtrees of the decision tree. This approach
%% is more efficient during the construction of the lower parts of the tree
%% where there is more parallelism \cite{DBLP:journals/pc/CoppolaV02}.

%% Finally, some other approaches try to combine at different tree level
%% construction the above two strategies \cite{SAR99,ZHA99}.

%SLIQ \cite{MAR96} has been among the first scalable algorithms for decision tree
%induction.
%The parallel implementation of the SLIQ classifier uses a
%memory-resident data structure which stores the class labels of each record.
%This in-core data structure impose serious memory requirements limiting
%the size of the datasets SLIQ can handle.
%A more memory-efficient version of SLIQ is SPRINT \cite{SAM96}.

Several works focus on distributed-memory machines, including SPRINT \cite{SAM96}, ScalParC
\cite{JKK98}, pCLOUDS \cite{SAR99}, and the approach of \cite{DBLP:journals/pc/CoppolaV02}. On
these shared-nothing architectures the cost of communications is the critical aspect. The use of
scalable data structures and of efficient load balancing techniques, trying to minimise costly data
redistribution operations, are the most important factors to obtain good performance. As an
example, the earliest SPRINT parallel algorithm adopts scalable SLIQ data structure for
representing the dataset, but it suffers from communication bottlenecks addressed in the successor
system ScalParC.
%SPRINT \cite{SAM96} is a memory-efficient parallel algorithm for decision tree induction
%which uses SLIQ scalable data structures.
%This algorithm partition the attribute lists in a row-block fashion among the
%processors ensuring that the continuous attributes maintain their sorted
%order on values at each nodes. Each processor determines the best split
%points for all the records it has, and using a all-to-all communication
%schema it is determined the best global split point.
%The result is used to partition the attribute lists of the splitting attribute.
%In order to split other lists, a hash table is used to maintain a mapping
%of record ids to nodes. The hash table needs to be communicated all the other
%processors from the processor that produced the winning split point.
%The other processors then use the hash table to split their own attribute lists.
%The size of this hash table is proportional to the total number of
%records in the training set.
%The algorithm uses an hash table, whose size is proportional to the total number
%of training cases. This table needs to be communicated all the other
%processors inducing communication bottlenecks during the computation.
% from the processor that produced the winning split point.
%% To overcome the problem of the hash table size, and the bottleneck produced
%% in the hash table communication, a distributed hash table implementation is
%% used in ScalParC \cite{JKK98}.
%To overcome these bottlenecks, %produced in the hash table communication,
%a distributed hash table implementation is used in ScalParC \cite{JKK98}.
pCLOUDS \cite{SAR99} combines both the data parallel  and the task parallel approaches. It exploits
data  parallelism for large decision nodes, then it switches to a task parallel approach as soon as
the nodes become small enough.
%The assigning and processing of small nodes are delayed
%until all the large nodes have been processed to reduce the number of message
%startups.
%Differently from SLIQ, SPRINT and ScalParC, pCLOUDS does not pre-sort the
%attribute lists, but uses sampling for deriving the best splitting
%points.
%, i.e. evaluates only a subset of the splitting
%points, for deriving the best splitting points.
The proposal of \cite{DBLP:journals/pc/CoppolaV02} categorises tasks in three different classes:
large, intermediate and small ones. Large tasks process a decision node, as to increase the degree
of available parallelism. Intermediate tasks process a sub-tree up to a given number of nodes and
within computation time bounds. Small tasks sequentially process the whole sub-tree of a node.

\yadtff takes inspiration from the two latter works and distinguish from them for: 1) it does not
need the redesign of the sequential algorithm but rather an \emph{easy-yet-efficient} porting of
the existing code; 2) it targets multicore rather than distributed memory machines; 3) it adopts an
effective cost model for deciding whether to parallelise on nodes or on attributes.
%Both in pCLOUDS and in the work presented in \cite{DBLP:journals/pc/CoppolaV02},
%it is not presented any analytical model for switching from data parallelism
%to task parallelism.
%RainForest is a general framework for scaling decision tree
%construction process to disk-resident data sets \cite{GRG00}.
%A parallelization for shared-memory multiprocessors
%of the RainForest RF-read algorithm has been proposed in \cite{JYA05}.
%The authors pointed out that the techniques typically used for
%parallelizing association mining and clustering can also be effectively
%used in parallelizing decision tree construction.
In conclusion, to the best of our knowledge, very few works specifically target data mining systems
on multicore architectures \cite{JYA05,GBPKNCD05,Bphd08,PZOC06}, but none specifically decision
tree algorithms.
%% In \cite{GBPKNCD05} it is proposed a tree structure that helps in improving benefits
%% of pre-fetching and a novel data structure that helps to achieve
%% high temporal locality for frequent pattern mining algorithms.
%% In \cite{Bphd08} it is proposed a parallel graph mining technique that obtain very
%% good speedup for chip-multi-processors architectures.
%A good survey that categorises  approaches to accelerate data
%mining according the different optimisation techniques the can be found in  \cite{PZOC06}.

\begin{table}[tb]
\label{fig:seqtimetable}
\begin{center}
%\parbox{0.8\linewidth}{\hspace{2ex}
\textsf{\scriptsize
\begin{tabular}[b]{l@{\hspace{4ex}}r@{\hspace{2ex}}r@{\hspace{2ex}}r@{\hspace{2ex}}r@{\hspace{5ex}}r}
\toprule
%&&\multicolumn{3}{c}{\centering FastFlow threads} & \\
&& 1E+1W & 1E+2W & 1E+3W &\\
\cmidrule(r){3-5}
{${\cal T}$ name} & Seq.Time (S) & \multicolumn{3}{c}{\centering Time (S)} & Max Boost \\
\cmidrule{1-6}
{\em Census PUMS}  & 4.46   & 4.3 & 2.37 & 1.69 &    $2.64\times$   \\
{\em U.S. Census}  & 17.67  & 17.97 & 11.17 & 7.8 &  $2.26\times$   \\
{\em KDD Cup 99}   & 18.11  & 17.26 & 9.12 & 6.67 &  $2.71\times$   \\
{\em Forest Cover} & 16.99  & 16.97 & 8.74 & 5.86 &  $2.90\times$   \\
{\em SyD10M9A}     & 103.21 & 93.95 & 52.34 & 39.37& $2.62\times$   \\
\bottomrule
\end{tabular}
}
\end{center}
%\vspace{-2ex}
\caption{ \yadt vs \yadtff on a Nehalem quad-core (E= Emitter, W=Worker).
%Test executed on a quad-core Xeon E5420@2.5GHz.
\label{tab:seqtime}} \vspace{-1cm} \end{table}

\section{Conclusions}

Nowadays, and for foreseeable future, the performance improvement of a
single core will no longer satisfy ever increasing computing power demand.
%follow the Moore's law.
For this, computer hardware industry shifted to multicore, and thus the extreme optimisation of
sequential algorithms is not longer sufficient  to squeeze the real machine power. Software
designers
%including data miners,
are then required to develop and to port applications on multicore. In this paper, we have
presented the case study of decision tree algorithms, by
porting \yadt using the \ff parallel programming framework. The strength of our approach consists
in the minimal change of the original code with, a the same time, a non-trivial parallelisation
strategy implementation (nodes and attributes parallelism plus
weighted problem-aware load balancing) and
with notable speedup. Eventually, we want to stress the results in the case of a low
cost quad-core architecture that may  be currently present in the desktop PC of any data analyst.
Table~\ref{tab:seqtime} shows that the parallelisation of \yadt boosts up to $2.9 \times$, with no
additional cost to buy a specific parallel hardware.

% As further work extension, we think that the SMT effectiveness for the \yadtff
% can be further exploited by designing a cache-aware weighted scheduling
% policy. Furthermore, with minor modifications to the \ff run-time, it is
% possible to improve the overall efficiency of the proposed approach by letting
% the Emitter thread to act as a Worker thread during idle times. We believe that
% the proposed methodology can also be applied to other mining algorithm.

%\bibliographystyle{abbrv}
%\bibliography{pkdd,ind,UniPisaGroup,ac,grid,multicore}

\section*{Acknowledgements}
The final, revised publication is available at \url{www.springerlink.com}\\ 

\noindent
M.~Aldinucci, S.~Ruggieri, and M.~Torquati. Porting decision tree
algorithms to multicore using FastFlow. In J.~L.~Balc{\'a}zar, F.~Bonchi,
A.~Gionis, and M.~Sebag, editors, {\em Proc. of European Conference in
  Machine Learning and Knowledge Discovery in Databases (ECML PKDD)},
volume 6321 of LNCS, pages 7--23, Barcelona, Spain,
Sept. 2010. Springer. DOI: \url{10.1007/978-3-642-15880-3_7}

%at \url{http://dx.doi.org/10.1007/978-3-642-15880-3_7}

\end{document}